\documentclass[12pt]{article}

\usepackage{multirow}
\usepackage{amsmath,amssymb,graphicx, graphics, color} 
\usepackage{bbm}
\usepackage{cite}
\usepackage{latexsym}
\usepackage{cancel}
\usepackage{enumerate}

\usepackage[normalem]{ulem} 

\usepackage{dsfont}

\newcommand{\centeron}[2]{{\setbox0=\hbox{#1}\setbox1=\hbox{#2}\ifdim
\wd1>\wd0\kern.5\wd1\kern-.5\wd0\fi \copy0
\kern-.5\wd0\kern-.5\wd1\copy1\ifdim\wd0>\wd1
                                   \kern.5\wd0\kern-.5\wd1\fi}}
\newcommand{\ltap}{\>\centeron{\raise.35ex\hbox{$<$}}
                           {\lower.65ex\hbox{$\sim$}}\>}
\newcommand{\gtap}{\>\centeron{\raise.35ex\hbox{$>$}}
                           {\lower.65ex\hbox{$\sim$}}\>}

\newcommand\ZZ{\hbox{\zfont Z\kern-.4emZ}}
\font\zfont = cmss10 

\newcommand{\eref}[1]{Eq.\ (\ref{e.#1})}

\newcommand{\sref}[1]{section~\ref{s.#1}}

\newcommand{\cref}[1]{Chapter \ref{c.#1}}

\newcommand{\ba}{\begin{array}}
\newcommand{\ea}{\end{array}}

\newcommand{\beq}{\begin{eqnarray}}
\newcommand{\eeq}{\end{eqnarray}}
\newcommand{\beqs}{\begin{eqnarray*}}
\newcommand{\eeqs}{\end{eqnarray*}}

\newcommand{\bal}{\begin{align}} 
\newcommand{\eal}{\end{align}}

\def\bi{\begin{itemize}}
\def\ei{\end{itemize}}
\def\ben{\begin{enumerate}}
\def\een{\end{enumerate}}
\def\bc{\begin{center}}
\def\ec{\end{center}}
\def\bt{\begin{table}}
\def\et{\end{table}}
\def\btb{\begin{tabular}}
\def\etb{\end{tabular}}

\def\mass2{mass${}^2$}

\newcommand{\tr}{\mathrm{Tr}}

\def\MB{\mathcal{M}}

\begin{document}
\bibliographystyle{unsrt}
\begin{titlepage}

\vskip2.5cm
\begin{center}
\vspace*{5mm}
{\huge \bf Spontaneous $R$-symmetry Breaking }
\vskip3mm
{\huge \bf with Multiple Pseudomoduli}
\end{center}
\vskip0.2cm

\begin{center}
{\bf David Curtin,$^{1}$ Zohar Komargodski,$^{2,3}$ David Shih,$^{4}$ Yuhsin Tsai$^{5}$}

\end{center}
\vskip 8pt

\begin{center}
{\it $^1$ YITP, Stony Brook University, Stony Brook, NY 11794.}\\
\vspace*{0.3cm}
{\it $^2$ Weizmann Institute of Science, Rehovot 76100, Israel.}\\
\vspace*{0.3cm}
{\it $^3$ Institute for Advanced Study, Princeton, NJ 08540, USA.}\\
\vspace*{0.3cm}
{\it $^4$ NHETC, Rutgers University, Piscataway, NJ 08854.} \\
\vspace*{0.3cm}
{\it $^5$Department of Physics, LEPP, Cornell University, Ithaca, NY 14853.} \\
\vspace*{0.3cm}

\end{center}

\vglue 0.3truecm

\begin{abstract}
We examine generalized O'Raifeartaigh models that feature multiple tree-level flat directions and only contain fields with $R$-charges 0 or 2.  We show that spontaneous $R$-breaking at up to one-loop order is impossible in such theories. Specifically, we prove that the  $R$-symmetric origin of field space is always a local minimum of the one-loop Coleman-Weinberg potential,  generalizing an earlier result for the case of a single flat direction. This result has consequences for phenomenology and helps elucidate the behavior of various models of dynamical SUSY breaking.
\end{abstract}

\end{titlepage}

\section{Introduction}
\label{s.intro} \setcounter{equation}{0} \setcounter{footnote}{0}

The O'Raifeartaigh (O'R) model~\cite{ORorig} and its generalizations constitute the simplest theories which spontaneously break supersymmetry (SUSY). Despite their simplicity, they are interesting subjects for study, because they often arise as low-energy effective theories of models which dynamically break SUSY~\cite{ISS, retrofitting, ITIY}. 

Spontaneous SUSY breaking generically requires the existence of an $R$-symmetry~\cite{NelsonSeiberg}. An unbroken $R$-symmetry forbids Majorana gaugino masses, so if SUSY is relevant to nature at the TeV scale, $R$-symmetry must be broken somehow.\footnote{One can also consider models with Dirac mass terms for the gauginos, see for instance the general analysis of~\cite{Benakli:2008pg} and references therein.}  In this paper we will examine the possibility of spontaneous $R$-symmetry breaking in generalized O'Raifeartaigh models (renormalizable Wess-Zumino models with $F$-term SUSY-breaking).

In general one can envision either tree-level spontaneous $R$-breaking or radiatively induced breaking~\cite{VCW}. Models that break the $R$-symmetry at tree-level exist~\cite{Carpenter, Sun, KomargodskiShih}, but they are rather cumbersome and have not yet been found naturally in dynamical models of SUSY breaking. One is therefore led to investigate the possibility of radiatively broken $R$-symmetry. In fact, radiative effects in Wess-Zumino models have always played a pivotal role because any SUSY-breaking vacuum is necessarily accompanied by a flat direction~\cite{ORhavePMS}. (Such flat directions in Wess-Zumino models are often called pseudomoduli.) Hence, to determine the correct vacuum of the theory one is generally forced to consider radiative effects.

A special class of generalized O'R models consists of theories where all the $R$-charges are either $0$ or $2$. Several well-known dynamical models of calculable SUSY breaking lead to such theories (e.g.\ \cite{ISS,ITIY}); hence our interest in this class.
For such theories one can prove the absence of tree-level $R$-breaking~\cite{KomargodskiShih}. In addition, it was shown in 
\cite{shihproof} that in models with a {\it single} pseudomodulus, spontaneous $R$-breaking through the one-loop Coleman-Weinberg (CW) potential required the presence of fields with $R$-charges other than 0 or 2. This theorem has often been used to guide model building.

In this paper, we will generalize the result of \cite{shihproof} to O'R models with arbitrarily many  pseudomoduli fields. We will show that even in this case, if all the $R$-charges are 0 or 2, the one-loop effective potential has a local minimum at the $R$-symmetric origin of field space (which could be a manifold in general). 
Additionally, we will also show that pseudomoduli can remain massless after one-loop corrections are taken into account only if they are in fact manifestly decoupled in the Lagrangian at the one-loop level. Such pseudomoduli can receive important two-loop corrections (see e.g.\ \cite{multiloopRbreaking1, multiloopRbreaking2, multiloopRbreaking3}), and it would be interesting to investigate these two-loop corrections in general. 

We do not consider the general problem of Wess-Zumino models with $R$-charges other than $0,2$. That is left as an interesting problem for the future. For the case of a single pseudomodulus it was  argued~\cite{shihproof} that there is no obstruction to obtaining $R$-symmetry breaking. It would be interesting to see precisely how this works if more than one pseudomodulus is present. Another obvious generalization of our study is to introduce gauge fields. Introducing gauge fields can lead to a variety of interesting phenomena, such as spontaneous radiative breaking, and even classical destabilization of all the vacua~\cite{Matos}. 

This work was partly motivated by recent interesting papers which considered the possibility of spontaneous $R$-symmetry breaking with additional pseudomoduli~\cite{Shadmi, Evans}.
In specific models, it was found by explicit computations that loop corrections preserve the $R$-symmetry.   In \cite{Evans}, it was also shown that having a \emph{single} additional pseudomodulus did not induce spontaneous $R$-breaking at one-loop and at leading order in SUSY breaking. Here we provide the general derivation for arbitrarily many pseudomoduli, and to all orders in SUSY breaking.

Our short note proceeds as follows. In \sref{modeldef} we define  the most general O'Raifeartaigh model containing only $R$-charges 0 and 2 and discuss the relevant terms. In  \sref{PMmassoneloop} we explicitly calculate the effective potential, and show that the generated mass matrix for $R=2$ fields at the origin is positive semi-definite. For completeness, we analyze the zero modes of this mass matrix in \sref{vanishingoneloop} and explicitly confirm that they can only arise for fields that are completely decoupled from SUSY breaking at one-loop order. For such zero modes one would need to investigate higher-order effects in order to determine the vacuum of the theory (or its absence). An appendix summarizes some technical details pertaining to \sref{vanishingoneloop}.

\section{Model Definition}
\label{s.modeldef} \setcounter{equation}{0}

Consider any theory with $R$-charges $0$, $2$  only. Label the $R$-charged fields $X$, $\sigma_i$, $i=1,\dots,N_2$ and the $R=0$ fields $\rho_a$, $a=1,\dots,N_0$. 
By a simple scaling argument of the $R$-charged fields, it is clear that from any field configuration one can find a path that terminates at $\sigma_i=0$ and along which the tree-level potential strictly decreases~\cite{KomargodskiShih}. In other words, from every point one can continuously lower the classical energy until an $R$-symmetric point is reached. (It can also be that the energy along this path stays constant, but by simply rescaling the $R$-charged fields it can never grow.) This makes tree-level breaking of the $R$-symmetry in such models impossible, and one has to rely on radiative corrections.  

Consider now the most general O'R model containing only $R$-charges 0 and 2.
Then the superpotential can always be brought into the canonical form  \cite{ORhavePMS,KomargodskiShih}
\begin{equation}
\label{e.canonicalform}
W = f X + m_{ai} \rho_a \sigma_i + \lambda_{ab} X \rho_a \rho_b + \tilde\lambda_{iab} \sigma_i \rho_a \rho_b,
\end{equation}
where $R(X) = 2$ and $\rho$, $\sigma$ are as above. $X$ is the canonical SUSY-breaking pseudomodulus. There can also be other pseudomoduli not associated with SUSY breaking. If these have $R=0$, then we do not care what happens to them radiatively, since they will not break $R$-symmetry regardless. Therefore we are free to expand those $\rho$ fields which are pseudomoduli around their exact vevs.\footnote{Radiatively-generated SUSY-breaking tadpoles in the scalar potential will shift the classical vevs for all the $R=0$ fields away from the origin if they are not protected by additional symmetries, but as long as those corrections are small we need not worry about them.} On the other hand, any additional $R=2$ pseudomoduli are potentially important. If they get vevs radiatively then they will break $R$-symmetry spontaneously.  So our task is to compute the Coleman-Weinberg potential in this multi-dimensional space and show that the $R$-symmetric origin is attractive.

There are additional $R=2$ pseudomoduli if and only if ${\rm rank}\, \,m < N_2$. Let us single out those that do not have mass terms and call them $Y_n$, $n=1,\dots,N_{2}'$. We will continue to denote the massive $R=2$ fields with $\sigma_i$, with an obvious reduction in their number. Then we can rewrite (\ref{e.canonicalform}) as
\begin{equation}
\label{e.canonicalform2}
W = f X + m_{ai} \rho_a \sigma_i + \lambda_{ab} X \rho_a \rho_b +\tilde\lambda_{nab} Y_n \rho_a \rho_b + \tilde\lambda_{iab}' \sigma_i \rho_a \rho_b,
\end{equation}
where $m^\dagger m$ is non-singular. Note that $m m^\dagger$ could have zero modes, but there are no tree-level tachyons at the origin. For the purposes of computing the one-loop effective potential for $X$ and $Y$, the cubic couplings $\tilde\lambda'$ never contribute, so we will ignore them henceforth and focus on the simplified superpotential
\begin{equation}
\label{e.canonicalform3}
W = f X + m_{ai} \rho_a \sigma_i + \lambda_{ab} X \rho_a \rho_b  +\tilde\lambda_{nab} Y_n \rho_a \rho_b,
\end{equation}
Finally, it is convenient to introduce a pseudomoduli-dependent matrix $N_{ab}$ defined by
\begin{equation}
\label{e.Ndef}
N_{ab} = \lambda_{ab} X + \tilde\lambda_{nab}Y_n.
\end{equation}
Note that $N$ can be taken to be symmetric (but not necessarily real) without loss of generality. In the next section, we will compute the Coleman-Weinberg one-loop effective potential for $X$ and $Y_n$ that follows from this superpotential.

\section{Pseudomoduli Masses at 1-Loop}
\label{s.PMmassoneloop} \setcounter{equation}{0} 
In terms of the tree-level boson and fermion mass matrices, the 1-loop effective potential  \cite{VCW} is given by
\begin{equation}
\label{e.Veff}
V_{eff}^{(1)} = \frac{1}{64 \pi^2} \sum_{i = F, B} \mathrm{Tr}(-1)^F \MB_i^4 \log \frac{\MB_i^2}{\Lambda^2}.
\end{equation}
Following~\cite{shihproof}, we rewrite this as
\begin{equation}
\label{e.Veffii}
V_{eff}^{(1)} = - \frac{1}{32 \pi^2}\int_0^\Lambda dv \,v^5\, \tr  \left( \frac{1}{v^2 + M_B^2} - \frac{1}{v^2 + M_F^2}\right).
\end{equation}
The mass matrices that follow from (\ref{e.canonicalform2}) are (in the basis $(\rho,\sigma,\rho^*,\sigma^*)$)
\begin{align}
\label{e.mb1}
M_B^2 =& \left( 
\begin{array}{cc} 
W_{ik}^\dagger W^{kj} & W_{ijk}^\dagger W^k\\
W^{ijk} W_k^\dagger & W^{ik} W_{kj}^\dagger \end{array}\right)  = M_0^2+M_1^2+M_2^2+ F \displaybreak[1] \\
\label{e.mb2}
M_0^2 =& \left(\begin{array}{cccc} m^* m^T & 0 & 0 & 0 \cr 0 & m^\dagger m & 0  &0  \cr 0 & 0  & m m^\dagger &  0 \cr 0 & 0 & 0 & m^T m^*\end{array}\right) \displaybreak[1] \\
\label{e.mb3}
M_1^2 =& \left(\begin{array}{cccc} 0 & N^\dagger m & 0 &  0 \cr m^\dagger N & 0 & 0  & 0 \cr 0 & 0 & 0 & N^T m^*\cr 0 & 0 & m^T N^* & 0\end{array}\right) \displaybreak[1] \\
\label{e.mb4}
M_2^2 =& \left(\begin{array}{cccc} N^\dagger N & 0 & 0 & 0 \cr 0 & 0 & 0 & 0  \cr  0 & 0 & N^T N^*  &0 \cr 0 & 0 & 0 & 0 \end{array}\right) \displaybreak[1]  \\
\label{e.mb5}
F =& \left(\begin{array}{cccc}0 & 0 & \lambda^\dagger f & 0 \cr 0 & 0 & 0  &0  \cr \lambda f^* & 0  & 0 &  0 \cr 0 & 0 & 0 & 0\end{array}\right)
\end{align}
and $M_F^2$ is the same but with $F\to 0$.  We would like to expand (\ref{e.Veffii}) out to second order in $N$. Using (\ref{e.mb1})-(\ref{e.mb5}), we obtain
\begin{eqnarray}
\nonumber V_{eff}^{(1)}\Big|_{N^2} & = & {1\over32\pi^2}\int_0^\Lambda dv\, v^5\,{\rm Tr}\, \Bigg( (v^2+M_0^2+F)^{-2}(M_2^2-M_1^2 (v^2+M_0^2+F)^{-1}M_1^2 ) - (f\to 0)\Bigg) \cr
 & =  & {1\over16\pi^2}\int_0^\infty dv\, v^3\,{\rm Tr}\, \Bigg( (v^2+M_0^2+F)^{-1}\left(M_2^2-{1\over2}M_1^2 (v^2+M_0^2+F)^{-1}M_1^2 \right) - (f\to 0)\Bigg) \\
  &&
\end{eqnarray}
where in the second line we have integrated by parts. This is the generalization of Eqn. (2.12) in \cite{shihproof}.

Next we expand out $(v^2+M_0^2+F)^{-1}$ in powers of $F$, delete the terms that vanish under the trace, and resum the series. This results in:
\begin{equation}
V_{eff}^{(1)}\Big|_{N^2}  = {1\over16\pi^2}\int_0^\infty dv\, v^3\,{\rm Tr}\, \left({\hat F^2\over 1-\hat F^2}\left(\hat M_2^2 - \hat M_1^4 \right)\right),
\end{equation}
where  the hatted quantities are defined by
\begin{eqnarray}
\hat F &=& (v^2+M_0^2)^{-1/2}F(v^2+M_0^2)^{-1/2},
\\
\hat M_{1,2}^2 &=& (v^2 + M_0^2)^{-1/2} M_{1,2}^2 (v^2 + M_0^2)^{-1/2}.
\end{eqnarray}
(Since $M_0$ can be singular this may not be well-defined at $v = 0$, but this does not matter for the $v$-integral.) Evaluating the block-matrix multiplication and making use of the fact that $\lambda, \tilde \lambda_n$ are symmetric, this finally becomes
\begin{equation}
\label{e.pmmsqfinal}
V_{eff}^{(1)}\Big|_{N^2}  = {1\over8\pi^2}\int_0^\infty dv\, v^5 \,{\rm Tr}\,\left({\hat\lambda^\dagger\hat\lambda \over 1-\hat\lambda^\dagger\hat\lambda}\hat N^\dagger \hat N\right),
\end{equation}
where
\begin{eqnarray}
\label{e.hatNlambda}
\hat\lambda &\equiv& (v^2+mm^\dagger)^{-1/2}\lambda f^*(v^2+m^*m^T)^{-1/2},\nonumber\\
\hat N &\equiv& (v^2+mm^\dagger)^{-1/2} N (v^2+m^*m^T)^{-1/2}.
\end{eqnarray}
The absence of tree-level tachyons at the origin implies that $m m^\dagger$ is positive-semidefinite. Therefore $(1-\hat \lambda^\dagger \hat \lambda)^{-1}$ is positive-semidefinite, which makes  the integrand a trace of a product of positive-semidefinite Hermitian matrices. Hence it is manifestly non-negative for all $X$ and $Y_n$, making all pseudomoduli masses non-tachyonic at the origin. Generally, they will have positive mass-squareds; we will examine the case where their mass-squareds vanish in the next section.

\section{Vanishing 1-Loop Masses}
\label{s.vanishingoneloop} \setcounter{equation}{0}

We have so far shown that the pseudomoduli mass-squareds around the origin are all non-negative, and thus there is no $R$-breaking at one-loop in the sense defined before. 
To complete the story we need to discuss the pseudomoduli which are massless at one-loop. We will show that this is only possible if these pseudomoduli are manifestly decoupled from SUSY breaking at one-loop order. This shows that there are no possible accidental cancellations, and all the pseudomoduli that can become massive indeed do so. Pseudomoduli which are manifestly decoupled at one-loop can still communicate with SUSY breaking at two and higher loops, and there are known examples where two-loop effects trigger spontaneous 
$R$-breaking~\cite{multiloopRbreaking1,multiloopRbreaking2,multiloopRbreaking3}. It would be interesting to say something general about the two-loop effective potential, but this is beyond the scope of this note.

In terms of the superpotential (\ref{e.canonicalform2}), what we would like to show is that if some pseudomodulus direction\footnote{Here we are being careful to distinguish between the pseudomodulus vevs $X_0$, $Y_{n0}$, and their fluctuations $\delta X\equiv X-X_0$, $\delta Y_n\equiv Y-Y_{n0}$.},
labelled  by $N_{ab}=   \lambda_{ab} X_0(t) + \sum_n Y_{n0}(t) \tilde\lambda_{nab}$ with $t \in \mathbb{R}$,
 is massless at one-loop, then $\rho$ and $\sigma$ can be split into two nearly-decoupled sets of fields $\{\rho\} \to\{\rho',\rho''\}$, $\{\sigma\}\to \{\sigma',\sigma''\}$:
\begin{equation}\label{e.canonicalformdec}
W =\Big( f \delta X + \rho'^T m' \sigma' + \delta X \rho'^T \lambda\rho' \Big) + \Big( \rho''^T m'' \sigma''   + \rho''^T N\rho''\Big) + {\rm cubic}
\end{equation}
These fields only talk to each other through the cubic interactions (which include terms like $\sigma\rho\rho$ and $\delta Y \rho \rho$), and so the pseudomoduli $N$ acquire SUSY-breaking masses only at two and higher loops.

We will take a constructive approach to deriving (\ref{e.canonicalformdec}). 
That is, we will start from the formula for the one-loop pseudomoduli mass-squareds (\ref{e.pmmsqfinal}), use this to derive constraints on $\lambda$, $N$, and $m$ in the superpotential (\ref{e.canonicalform2}) in the event that the mass-squareds vanish, and show that these constraints necessarily lead us to the nearly-decoupled form (\ref{e.canonicalformdec}). 

To begin, suppose the mass of some pseudomodulus vanishes at one-loop order. According to (\ref{e.pmmsqfinal}), this means that
\begin{equation}
\label{e.msqvan}
\mathrm{Tr}(\hat \lambda^\dagger \hat \lambda \hat N^\dagger \hat N) = 0
\end{equation}
for $N$ in the background field direction of this zero mode. This in turn can only be satisfied if
\begin{equation}
\label{e.hatlNreq}
\hat \lambda \hat N^\dagger = 0
\end{equation}
Note that $\hat\lambda$ and $\hat N^\dagger$ are functions of $v$ via (\ref{e.hatNlambda}), and (\ref{e.hatlNreq}) must be true for all $v$. Expanding in $\frac{1}{v^2}$ yields the following conditions that must be satisfied by the coupling matrices:
\begin{equation}
\label{e.maincondition}
\lambda (m^*m^T)^k N^\dagger = 0 \ \ \  \ \ \ \ \mathrm{for\,\, all} \ \ \ \ k = 0, 1, 2, \ldots
\end{equation}
$\lambda$ is a complex symmetric matrix, so by a unitary rotation of the $\rho$ fields $\lambda\to U \lambda U^T$, we can always diagonalize it:
\begin{equation}
\label{e.lambdahatcond}
\lambda = \left( \begin{array}{cc} \lambda'_{n_1\times n_1} & 0\\  0 & 0 \end{array}\right)
\end{equation}
where $\lambda'$ is non-singular. The $k=0$ version of (\ref{e.maincondition}) implies $\lambda N^\dagger=0$, so in the basis where $\lambda$ takes the form (\ref{e.lambdahatcond}), we can do another unitary rotation on the $\rho$ fields not coupling to $\lambda'$ so that 
\begin{eqnarray}
\label{e.lambdaNcondition}
N &=& \left(\begin{array}{ccc} 0_{n_1 \times n_1} &  & \\  & 0_{n_2\times n_2} &  \\  &  & N'_{n_3 \times n_3} \end{array}\right),
\end{eqnarray}
with $N'$ non-singular.  $n_2 $ could of course be zero.

Having used the $k = 0$ condition of (\ref{e.maincondition}) to fix the block-form of $\lambda$ and $N$, the $k > 0$ conditions will restrict the form $m$. Writing the hermitian matrix $m^*m^T$ in $3 \times 3$ block form as in (\ref{e.lambdaNcondition}), the $k>0$ conditions of (\ref{e.maincondition}) imply
\begin{equation}
\label{e.mmdaggercond}
 ((m^*m^T)^k)_{13} = 0  \ \ \  \ \ \ \ \mathrm{for\,\, all} \ \ \ \ k = 1, 2, \ldots
\end{equation}
(The 13 subscript refers to the upper-right block of $(m^*m^{T})^k$.) In the appendix, we prove the following lemma in linear algebra: when  (\ref{e.mmdaggercond}) is satisfied, one can always find a $3\times 3$ block-unitary transformation that puts $m^* m^{T}$ into the form
\begin{equation}
m^* m^T = \left( \begin{array}{cc} \left[m^* m^T \right]'_{n_4\times n_4} & 0 \\ 0 & \left[m^* m^T\right]''_{n_5\times n_5} \end{array}\right)
\end{equation}
with $n_4\ge n_1$ and $n_5\ge n_3$. Combining this with (\ref{e.lambdahatcond}) and (\ref{e.lambdaNcondition}), we conclude that all the $\rho$ fields can be separated into two sectors in which $\lambda$, $N$, and $m^*m^T$ are block-diagonal. By a unitary transformation on the $\sigma$ fields, the same can be done for $m$ itself, and we arrive at the desired result (\ref{e.canonicalformdec}).

\section*{Acknowledgements}

We are grateful to Y.~Shadmi for helpful discussions. The work of D.C. was supported in part by the National Science Foundation under Grant PHY-0969739. Z.K. is supported by NSF PHY-0969448,  by a research grant from Peter and Patricia Gruber Awards, and by the Israel Science Foundation (grant $\#$884/11). The work of D.S. was supported in part by a DOE Early Career Award. In addition, the research of Z.K. and D.S. was supported in part by Grant No 2010/629 from the United States-Israel Binational Science Foundation (BSF).  The work of Y.T. was supported in part by the National Science Foundation under Grant No.~PHY-0757868. Opinions, conclusions or recommendations arising out of supported research activities are those of the author or the grantee and should not be presented as implying that they are the views of the funding agencies.

\appendix

\section{Useful Lemma}
\label{a.lemma} \setcounter{equation}{0}

In this appendix, we will prove the following lemma described above in section 4.
\medskip
\medskip

{\underline {\it Lemma:}} Consider a square hermitian matrix $M$, divided into blocks
\begin{equation}
\label{e.Mblocks}
M = \left( \begin{array}{ccc} M_{11} & M_{12} & M_{13} \\ M_{12}^\dagger & M_{22} & M_{23}\\ M_{13}^\dagger & M_{23}^\dagger & M_{33} 
\end{array}\right)
\end{equation}
with $M_{ij}$ being $m_i\times m_j$. Suppose that $M$ satisfies:
\begin{equation}
\label{e.Msat}
(M^k)_{13} = 0 \quad {\rm for\,\, all}\quad k=1,2,\dots
\end{equation}
Then there exists a block unitary transformation $M\to U M U^\dagger$ with
\begin{equation}
\label{e.Udef}
U = \left( \begin{array}{ccc}  U_1 &&\\  & U_2 &\\  & & U_3
\end{array}\right)
\end{equation}
such that $M$ takes the block-diagonal form 
\begin{equation}
\label{e.Mblocksii}
M =\left( \begin{array}{cc} \tilde M_{11} & 0 \\ 0 & \tilde M_{22}
\end{array}\right)
\end{equation}
with the $12$ block that is zero in \eref{Mblocksii} containing the 13 block in the original basis.

\medskip\medskip

{\underline{\it Proof:}} We will prove this by induction, by starting with general $m_1$, $m_2$, $m_3$ and then reducing this to the same claim but with smaller $m_i$. The $k=1$ version of \eref{Msat} implies that $M_{13}=0$. The $k=2$ condition implies that $M_{12}M_{23}=0$. Combining this with a choice of $U_1$, $U_2$ and $U_3$, we can always simultaneously block-diagonalize $M_{12}$ and $M_{23}$:
\begin{equation}
\label{e.ABCblock}
M_{12} = \left( \begin{array}{ccc} (A)_{m_1'\times m_1'} &  0 & 0\\0  & 0 & 0\end{array}\right)\quad 
M_{23} = \left( \begin{array}{cc} 0 & 0 \cr 0 & 0 \\ 0 &  (B)_{m_3'\times m_3'}\end{array}\right)
\end{equation}
with $A$ and $B$ nonsingular, and $m_1'\leq m_1, m_2$, $m_3'\leq m_2, m_3$, and $m_1'+m_3'\leq m_2$. Dividing  $M_{22}$ into $3\times3$ blocks like \eref{Mblocks} with $m_i \rightarrow m'_i$, the $k\geq 3$ versions of \eref{Msat} imply 
\begin{equation}
\label{e.Msatii}
((M_{22})^\ell)_{1'3'} = 0\quad {\rm for\,\, all}\quad \ell=1,2,\dots
\end{equation}
So we see that \eref{Msat} maps on to an identical condition for the smaller matrix $M_{22}$. Moreover, examining the form of \eref{Mblocks}, after substituting \eref{ABCblock}, we find:
\begin{equation}
\label{e.Mblocksiii}
M = \left( \begin{array}{ccc} 
M_{11} & 
\left( \begin{array}{ccc}A&  0 & 0\\0  & 0 & 0\end{array}\right)
 & 0
 \\
 \left(\begin{array}{cc} A^\dagger & 0 \\ 0 & 0 \\ 0 & 0 \end{array}\right) & M_{22} & 
 \left(\begin{array}{cc} 0& 0 \\ 0 & 0 \\ 0 & B \end{array}\right) 
 \\ 0 & 
 \left( \begin{array}{ccc}0&  0 & 0\\0  & 0 & B^\dagger\end{array}\right)
  & M_{33} 
  \end{array}\right)
\end{equation}                                  
So we see that the desired $2\times 2$ block form \eref{Mblocksii} can be achieved, provided $M_{22}$ can be put into an analogous $2\times 2$ block form, also with a block-unitary transformation.  This completes the inductive recursion.  Proceeding in this way, we can reduce the lemma to a trivial statement about $3\times 3$ matrices, which completes the proof by induction.


\vspace*{-3mm}
\begin{thebibliography}{99}


\bibitem{ORorig}
  L.~O'Raifeartaigh,
  ``Spontaneous Symmetry Breaking For Chiral Scalar Superfields,''
  Nucl.\ Phys.\ B {\bf 96}, 331 (1975).


  \bibitem{ITIY}	
  K.~A.~Intriligator and S.~D.~Thomas,
  Nucl.\ Phys.\ B {\bf 473}, 121 (1996)
  [hep-th/9603158];
    K.~-I.~Izawa and T.~Yanagida,
  Prog.\ Theor.\ Phys.\  {\bf 95}, 829 (1996)
  [hep-th/9602180].


\bibitem{ISS} 
  K.~A.~Intriligator, N.~Seiberg and D.~Shih,
  ``Dynamical SUSY breaking in meta-stable vacua,''
  JHEP {\bf 0604}, 021 (2006)
  [hep-th/0602239].


\bibitem{retrofitting}
  M.~Dine, J.~L.~Feng and E.~Silverstein,
  ``Retrofitting O'Raifeartaigh models with dynamical scales,''
  Phys.\ Rev.\ D {\bf 74}, 095012 (2006)
  [arXiv:hep-th/0608159].





\bibitem{NelsonSeiberg}
  A.~E.~Nelson and N.~Seiberg,
  ``R symmetry breaking versus supersymmetry breaking,''
  Nucl.\ Phys.\ B {\bf 416}, 46 (1994)
  [arXiv:hep-ph/9309299].


\bibitem{Benakli:2008pg} 
  K.~Benakli and M.~D.~Goodsell,
  ``Dirac Gauginos in General Gauge Mediation,''
  Nucl.\ Phys.\ B {\bf 816}, 185 (2009)
  [arXiv:0811.4409 [hep-ph]].


  \bibitem{VCW}
  S.~R.~Coleman, E.~J.~Weinberg,
  ``Radiative Corrections as the Origin of Spontaneous Symmetry Breaking,''
  Phys.\ Rev.\  {\bf D7 } (1973)  1888-1910.
  

  
\bibitem{Carpenter} 
  L.~M.~Carpenter, M.~Dine, G.~Festuccia and J.~D.~Mason,
  ``Implementing General Gauge Mediation,''
  Phys.\ Rev.\ D {\bf 79}, 035002 (2009)
  [arXiv:0805.2944 [hep-ph]].

\bibitem{Sun} 
  Z.~Sun,
  ``Tree level spontaneous $R$-symmetry breaking in O'Raifeartaigh models,''
  JHEP {\bf 0901}, 002 (2009)
  [arXiv:0810.0477 [hep-th]].

 \bibitem{KomargodskiShih} 
  Z.~Komargodski and D.~Shih,
  ``Notes on SUSY and $R$-Symmetry Breaking in Wess-Zumino Models,''
  JHEP\ {\bf 0904}, 093  (2009)
  [arXiv:0902.0030 [hep-th]].
  

\bibitem{ORhavePMS}
  S.~Ray,
  ``Some properties of meta-stable supersymmetry-breaking vacua in Wess-Zumino
  models,''
  Phys.\ Lett.\  B {\bf 642}, 137 (2006)
  [arXiv:hep-th/0607172].
  


\bibitem{shihproof}
  D.~Shih,
  ``Spontaneous $R$-symmetry breaking in O'Raifeartaigh models,''
  JHEP {\bf 0802}, 091 (2008).
  [arXiv:hep-th/0703196 [hep-th]].

  
  
  
 
  \bibitem{multiloopRbreaking1} 
  A.~Giveon, A.~Katz and Z.~Komargodski,
  ``On SQCD with massive and massless flavors,''
  JHEP\ {\bf 0806}, 003  (2008)
  [arXiv:0804.1805 [hep-th]].


\bibitem{multiloopRbreaking2} 
  K.~Intriligator, D.~Shih and M.~Sudano,
  ``Surveying Pseudomoduli: The Good, the Bad and the Incalculable,''
  JHEP\ {\bf 0903}, 106  (2009)
  [arXiv:0809.3981 [hep-th]].


\bibitem{multiloopRbreaking3} 
  A.~Amariti and A.~Mariotti,
  ``Two Loop $R$-Symmetry Breaking,''
  JHEP\ {\bf 0907}, 071  (2009)
  [arXiv:0812.3633 [hep-th]].



  
\bibitem{Matos} 
  L.~F.~Matos,
  ``Some examples of F and D-term SUSY-breaking models,''
  arXiv:0910.0451 [hep-ph].
  




%
\bibitem{Evans}
  J.~L.~Evans, M.~Ibe, M.~Sudano, T.~T.~Yanagida,
  ``Simplified $R$-Symmetry Breaking and Low-Scale Gauge Mediation,''
  [arXiv:1103.4549 [hep-ph]].
  
  \bibitem{Shadmi}
  Y.~Shadmi,
  ``Metastable Rank-Condition Supersymmetry Breaking in a Chiral Example,''
  JHEP {\bf 1108}, 149 (2011).
  [arXiv:1107.3565 [hep-th]].



\end{thebibliography}
\end{document}